# Astro2020 Science White Paper

# Kilonovae: nUV/Optical/IR Counterparts of Neutron Star Binary Mergers with *TSO*

**Thematic Areas:** ☐Planetary Systems ☐Star and Planet Formation
☒ Formation and Evolution of Compact Objects ☒ Cosmology and Fundamental Physics
☒ Stars and Stellar Evolution ☐Resolved Stellar Populations and their Environments
☒ Galaxy Evolution ☒ Multi-Messenger Astronomy and Astrophysics

**Principal Author:**
Name: Brian Metzger
Institution: Columbia University, Department of Physics
Email: bmetzger@phys.columbia.edu
Phone: 212-854-9702

**Co-authors:** (names and institutions)
Edo Berger[1], Jonathan Grindlay[1], Suvi Gezari[2], Zeljko Ivezic[3], Jacob Jencson[4], Mansi Kasliwal[4], Alexander Kutyrev[5], Chelsea Macleod[1], Gary Melnick[1], Bill Purcell[6], George Rieke[7], Yue Shen[8], Nial Tanvir[9], Michael Wood Vasey[10]

[1]Center for Astrophysics | Harvard and Smithsonian, USA, [2]University of Maryland, USA, [3]University of Washington, USA, [4]Caltech, USA, [5]NASA/GSFC, [6]Ball Aerospace, USA, [7]University of Arizona, USA, [8]University of Illinois, USA, [9]Leicester University, UK, [10]University of Pittsburgh, USA

**Abstract:** With the epochal first detection of gravitational waves from a binary neutron star (NS) merger with the GW170817 event, and its direct confirmation that NS-NS mergers are significant sources of the of the r-process nucleosynthesis of heavy elements, an immense new arena for prompt EM (X-rays through IR and radio) studies of fundamental physics has been opened. Over the next decade, GW observatories will expand in scale and sensitivity so the need for facilities that can provide prompt, high sensitivity, broad-band EM followup becomes more urgent. NS-NS or NS –black hole (BH) mergers will be instantly recognized (and announced) by the LIGO-international collaboration. LSST will be a prime resource for rapid tiling of what will usually be large ($\sim 10 - 100$ deg$^2$) error boxes. X-ray through IR Telescopes in space with (nearly) full-sky access that can rapidly image and tile are crucial for providing the earliest imaging and spectroscopic studies of the kilonova emission immediately following NS-NS mergers. The Time-domain Spectroscopic Observatory (*TSO*) is a proposed Probe-class 1.3m telescope at L2, with imaging and spectroscopy (R = 200, 1800) in 4 bands (0.3 - 5μm) and rapid slew capability to 90% of sky. *TSO* nUV-mid-IR spectra will enable new constraints on NS structure and nucleosynthesis.

**Neutron star mergers, Kilonovae and Nucleosynthesis: a new Era**

The merger of binary neutron stars, or of a neutron star and a stellar-mass black hole, results in the ejection of neutron-rich matter into space (e.g. Rosswog et al. 1999, Hotokezaka et al. 2013). As this material decompresses from nuclear densities, heavy nuclei are synthesized by the rapid capture of neutrons on lighter seed nuclei (r-process; Lattimer & Schramm 1974). The radioactive decay of these unstable isotopes powers an optical/infrared transient lasting from days to weeks or longer (Li & Paczynski 1998, Metzger et al. 2010).

The bolometric luminosity of the kilonova emission is primarily sensitive to the quantity of synthesized material, while the colors of the emission mainly probe the ejecta composition. In particular, whether the spectral energy distribution peaks at optical/UV wavelengths, or in the near-infrared, depends on the opacity of the ejecta, which in turn depends on the presence or absence, respectively, of heavy lanthanide nuclei (atomic mass number A> ~ 140; Kasen et al. 2013, Tanaka & Hotokezaka 2013). The time-evolution of the kilonova spectral energy distribution therefore contains key information on the radial and angular structure of the ejecta composition. For instance, while the tidal ejecta tends to be neutron-rich (red) and geometrically focused in the equatorial plane of the binary, outflows from the accretion disk can be less neutron-rich (blue and red) and more spherical in their geometry (e.g. Metzger & Fernandez 2014).

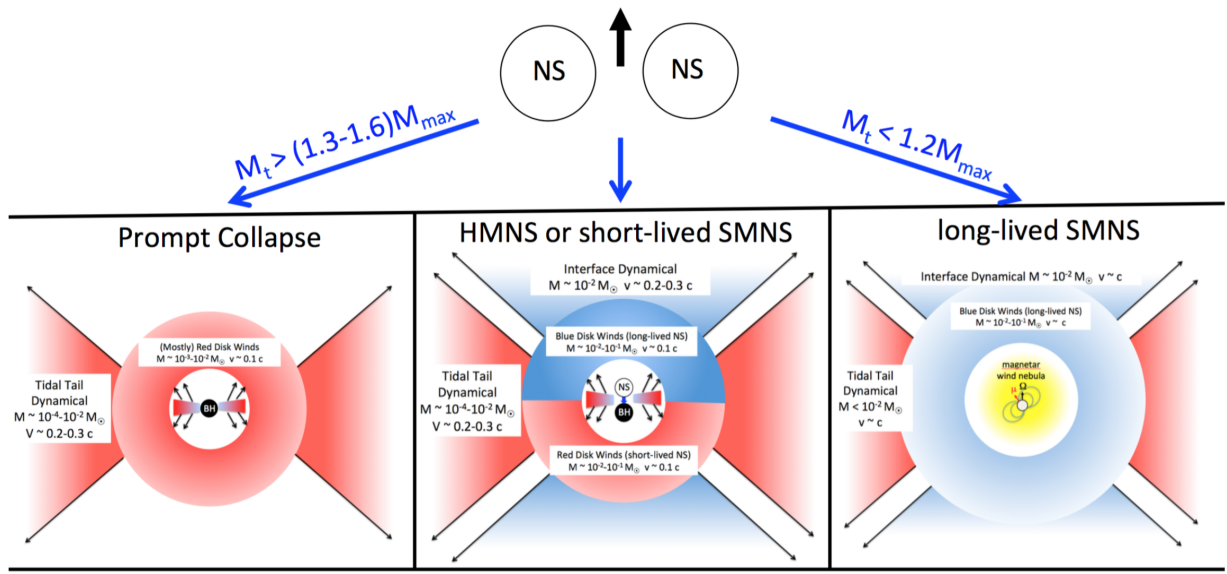

**Fig. 1.** The strength of the red and blue kilonova emission from a neutron star merger, and its geometry, depends on the compact remnant which forms immediately after the merger; the latter in turn depends on the total mass of the original binary or its remnant, $M_{tot}$, relative to the maximum NS mass, $M_{max}$. A massive binary results in a prompt collapse to a BH; in such cases, the polar shock-heated ejecta is negligible and the accretion disk outflows are weakly irradiated by neutrinos, resulting in a primarily red KN powered by the tidal ejecta (left panel). By contrast, a very low mass binary creates a long-lived supramassive NS, which imparts its large rotational energy to the surrounding ejecta (right panel). In the intermediate case, a hyper-massive or short-lived supra-massive NS forms, which produces both blue and red KN ejecta expanding at mildly relativistic velocities, consistent with observations of GW170817. From Margalit & Metzger (2017).

LIGO and Virgo are primarily sensitive to gravitational waves from the inspiral phase, but at their present sensitivities they cannot ascertain the merger outcome, such as the creation of a hyper-massive neutron star formation and its collapse into a black hole. Kilonovae thus provide unique probes the physics of the merger and its aftermath. The ejecta lanthanide fraction depends on the abundance of neutrons in the ejecta, which in turn depends on weak interaction processes that occur in the post merger environment (e.g. irradiation of the ejecta by neutrinos from the neutron star remnant prior to black hole formation). For instance, the large ejecta mass, and early blue colors of the kilonova observed following GW170817, was used to argue against a prompt collapse of the merger remnant to a black hole. Combining such inferences with information on the properties of the binary measured from the gravitational wave data, one can place constraints on the equation of state of dense nuclear matter (e.g. Margalit & Metzger 17, Bauswein et al. 2017). Future variations in the kilonova properties, for instance from particularly high or low mass binary neutron star systems, could be used to place even tighter constraints.

All double neutron merger mergers are expected to be accompanied by some mass ejection, making kilonova emission a ubiquitous signatures of these events. However, the relative strength of the optical versus infrared emission could vary between events, depending on the properties of the merging binary. For instance, for particularly massive NS binaries, the merger product immediately collapses into the black hole, reducing the mass of the accretion disk and thus the quantity of blue kilonova ejecta. Thus, in order to probe the full range of possible EM counterparts, and to connect them to information encoded in the gravitational wave data, well time-sampled infrared follow-up observations out to wavelengths of at least a few microns will be needed for a large sample of events.

Poorly located LIGO/VIRGO events could be rapidly tiled and located with LSST within a few hours. Long-term follow-up with both imaging and spectroscopy in the near-infrared is essential to follow the peak of the spectral energy distribution over timescales of weeks to months after the merger. This is precisely what *TSO* would enable. By measuring the bolometric light curve in the nebular phase through dedicated multi-band follow-up, we can ascertain details on the ejecta composition (Fig. 2) and could search for decay half-life signatures of individual radioactive isotopes (e.g. $^{254}$Cf), much in the same way that the $^{56}$Ni to $^{56}$Co chain is observed in normal supernovae (e.g. Zhu et al. 2018, Wanajo 2018, Wu et al. 2018). Late-

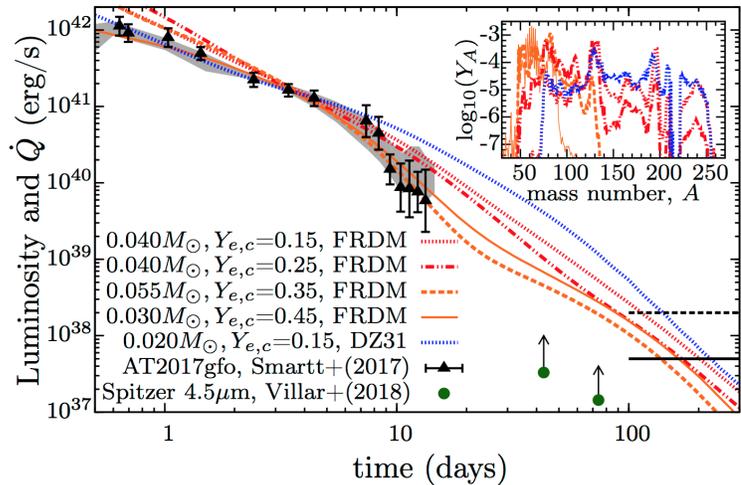

**Fig. 2**. Time sampled kilonova light curves, including into the late-time NIR nebular phase, are sensitive to the quantity and composition of the merger ejecta (e.g. electron fraction Ye) and may include the imprint of individual radioactive isotopes (from Wu et al. 2018).

time nebular observations are also sensitive to the lowest-velocity, innermost ejecta layers, which could produce emission line signatures indicative of individual elements.  UVOIR (Fig. 3)

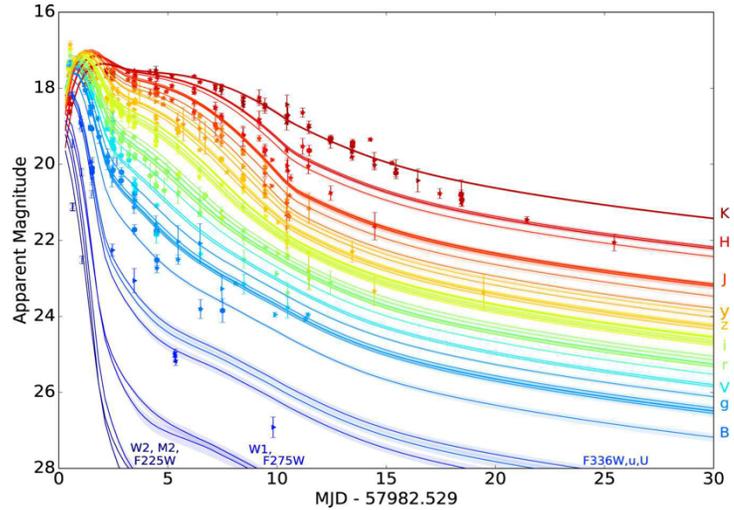

**Fig. 3.** UVOIR light curves of the kilonova associated with GW170817 from a compilation of published data, compared to two-component one-zone kilonova models, which include separate low-opacity ("blue") and high-opacity ("red") components (single-component models provide poor fits to the data). The best fit model has a total ejecta mass of ~5x10$^{-2}$ Msun and ejecta velocities ranging from 0.25 c (for the blue component) to 0.1 c (for the red component).  From Villar et al. (2017).

and IR (Fig. 4) lightcurves of GW170817 constrain the ejecta mass and velocity.

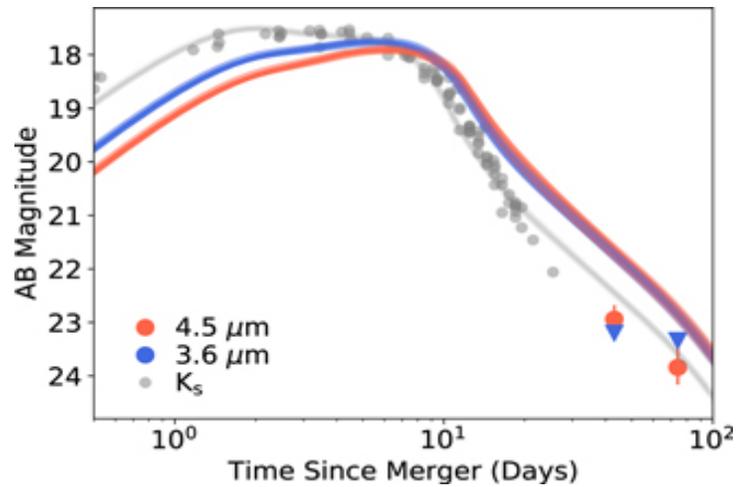

**Fig. 4.** nIR (Ks) and mid-IR (L, M) light curves of GW170817 Photometry and spectra from *TSO* would extend the coverage from the near UV to Mid-IR for all LIGO triggered NS mergers.
(from Villar et al. 2018; see also Kasliwal et.al. 2019)

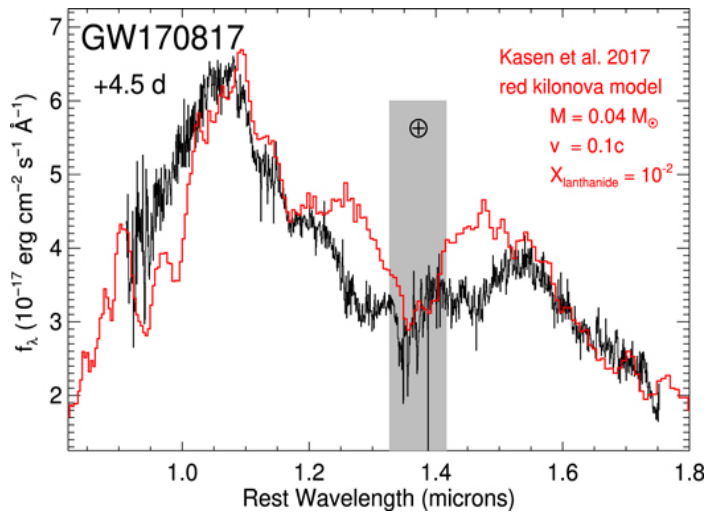

**Fig. 5.** infrared spectrum of GW170817 compared to current state-of-the-art models assuming an explosion velocity of v=0.1c. Although some features match reasonably well, there is clearly necessity to include additional physics to match others. (from Chornock et. al. 2017)

**BH-NS mergers**

In contrast to NS-NS mergers, whether significant mass ejection (and thus kilonova emission) is expected from NS-BH merger depends on whether the NS is tidally disrupted outside of the black hole event horizon (e.g. Foucart 2012). The latter condition depends on the mass and spin of the BH, as well as the radius of the NS, properties that again are probed in a complementary way by gravitational wave observations. A constraining non-detection of kilonova emission from a NS-BH merger could thus provide evidence for the presence of an event horizon. NS-BH mergers will be "louder" GW events, and better located, than NS-NS mergers. Prompt followup promptly with *TSO* to obtain $0.3 - 5.0\mu m$ spectra would provide information on the composition/velocity of the mass ejection mechanisms in NS-BH merger, which are key to placing constraints on the properties of the BH and the NS EOS.

**The ground-based GW detector network in the mid- to late-2020s**

The current GW detector "advanced" network consists of the two LIGO detectors in the US and the Virgo detector in Italy. In the 2nd science run the range for the LIGO detectors for binary neutron star mergers was ~80-100 Mpc. In the coming two science runs it is hoped to reach the original design sensitivity of LIGO/Virgo, increasing the range by a factor ~2, while at the same time the Japanese KAGRA instrument is expected to come on-line.

By the mid- to late-2020s further enhancements, in particular the so-called A+ development of LIGO, is expected to increase the range by a further factor of ~2, and IndIGO (the LIGO clone being constructed in India) will also come on-line resulting in improved source localisations over most of the sky (median ~10 sq-deg). Thus in this period the survey volume, and hence rate of detections, should be nearly two orders of magnitude greater than it has been to-date, with a corresponding increase in the follow-up challenge in terms of depth required and numbers of search campaigns that must be conducted. This challenge is one of the major motivations for *TSO*. Scientifically, this will provide the statistical samples to probe the diversity of compact binary merger behaviour, both of kilonovae and relativistic jets, allowing detailed tests of models, and exploring r-process yields as a function of system parameters.

**Counterpart identification**

A critical step in broad electromagnetic follow up is identification of a counterpart to the GW event, which requires wide-field imaging. This is most readily achievable in the optical band with a broad range of existing and upcoming facilities (particularly LSST). The example of GW170817 indicates that similar events (or even less luminous counterparts) will be detectable within the GW detector network horizon. In particular, rapid follow up with LSST will lead to detections of optical counterparts even for ejecta masses that are one-tenth of GW170817 (Cowperthwaite et al. 2018). These facilities will be capable of detecting counterparts regardless of the details of the nucleosynthesis. Once a counterpart is identified, multi-epoch spectroscopic and photometric follow-up is essential for tracking the evolution of the spectral energy distribution, spectral features and identification of potentially multiple ejecta components.

**Infrared Spectroscopy and Imaging**

Theory suggests that the range of kilonova behaviour is likely to be broad, and in particular cases of comparatively low ejecta masses will produce intrinsically faint events. This, combined with distances of potentially up to several hundred Mpc and a desire to monitor significantly after peak, necessitates sensitive spectroscopic follow-up. Similarly, there is a general expectation that the details of the nucleosynthesis (and potentially the viewing angle) will affect the observational signatures of a kilonova as a function of time. A robust prediction, borne out by observations of GW170817, is that within a few days post-merger the kilonova spectral energy distribution peaks in the infrared. It is therefore imperative to measure the spectra and light curves of kilonovae rapidly, across a broad wavelength range, and over a sufficiently long timescale to capture the full detailed nucleosynthesis and ejecta dynamics.

This is precisely what *TSO* would enable: the first simultaneous spectra in 4 bands from nUV to mid-IR (0.3 − 5.0μm), with resolution R = 200 for maximum sensitivity and IFU coverage of the kilonova and the surrounding array of 10 x 10 pixels (0.5") for host galaxy context. Deeper exposure R = 1800 spectra can also be obtained for events with light curve apparent magnitudes AB ≤23 (see *TSO*-WP, Grindlay et al). Given the GW170817 AB ~18 IR magnitudes (at 2.1, 3.6, 4.5μm) at T ~ 1 − 10 days after the outburst (cf. Fig. 5), and the ~40 Mpc distance of the host galaxy NGC 4993 (Hjorth +2017), in a 4 x $10^3$ sec (~1 hour) integration with the R = 200 IFU proposed for *TSO*, *the GW170817 event could be detected at AB ~23 with 10σ per pixel at a distance of 400 Mpc* ! The high resolution (R = 1800) grating would similarly detect this event at ~120 Mpc for the same exposure time. These sensitivities and band width are not possible from the ground, and the 0.3 − 0.6μm Blue coverage is not possible with JWST or WFIRST. It is clear that a space telescope is needed with rapid response to 90% of the sky, and with sensitivity, nUV to mid-IR band-width, and resolution similar to that proposed for the Probe-class *TSO* mission.